\documentstyle[emulateapj,psfig]{article}

\def\etal{{et~al.}}
\def\amin{\ifmmode^{\prime}\else$^{\prime}$\fi}
\def\asec{\ifmmode^{\prime\prime}\else$^{\prime\prime}$\fi}

\def\simgt{\lower.5ex\hbox{$\; \buildrel > \over \sim \;$}}
\def\simlt{\lower.5ex\hbox{$\; \buildrel < \over \sim \;$}}

\newcommand\xte{{\it RXTE}}

\newcommand\asca{{\it ASCA}}

\def\psr{\hbox{PSR~J1846$-$0258}}
\def\src{\hbox{AX~J184624.5$-$025820}}

\def\snr{\hbox{Kes~75}}

\lefthead{Gotthelf, Vasisht, Boylan-Kolchin, \&\ Torii}
\righthead{A 700 year-old Pulsar in Kes~75}

\begin{document}

\title{A 700 year-old Pulsar in the Supernova Remnant Kes~75}

\author{E. V. Gotthelf$^1$, G. Vasisht$^2$, M. Boylan-Kolchin$^1$, \&\ K.
Torii$^3$}
\altaffiltext{1}{Columbia Astrophysics Laboratory, Columbia University,
550 West 120$^{th}$ Street, New York, NY
10027, USA; evg@astro.columbia.edu; mbk@astro.columbia.edu}
\altaffiltext{2}{Jet Propulsion Laboratory,
California Institute of Technology, 4800 Oak Grove Drive, Pasadena,
CA, 91109, USA; gv@astro.caltech.edu}
\altaffiltext{3}{Space Utilization Research Program, Tsukuba Space Center, 
National Space Development Agency of Japan, 2-1-1 Sengen, Tsukuba, 
Ibaraki 305-8505, Japan; torii.kenichi@nasda.go.jp
}

\begin{abstract}
Since their discovery 30 years ago, pulsars have been understood to be
neutron stars (NSs) born rotating rapidly ($ \sim 10-100$ ms). These
neutron stars are thought to be created in supernova explosions
involving massive stars, which give rise to expanding supernova
remnants (SNRs).  With over 220 Galactic SNRs known (Green 1998) and
over 1200 radio pulsars detected (Camilo et al. 2000), it is quite
surprising that few associations between the two populations have been
identified with any certainty.  Here we report the discovery of a
remarkable 0.3 sec X-ray pulsar, \psr, associated with the supernova
remnant \snr.  With a characteristic age of only 723 yr, consistent
with the age of \snr, \psr\ is the youngest pulsar yet discovered
and is being rapidly spun down by torques from a large magnetic dipole
of strength $\simeq 5\times 10^{13}$ G, just above the so-called
quantum critical field.  \psr\ resides in this transitional regime
where the magnetic field is hypothesized to separate the regular
pulsars from the so-called magnetars. \psr\ is evidently a Crab-like
pulsar, however, its period, spin-down rate, spin-down conversion
efficiency, are each an order-of-magnitude greater, likely the result
of its extreme magnetic field.

\end{abstract}

\section {Introduction}

Kes 75 (also G29.7$-$0.3) is one of the few examples in our Galaxy of
a young, shell-type remnant ($3.5^{\prime}$ in diameter) with a
central core ($30^{\prime\prime}$) whose observed properties suggest a
synchrotron plerion similar to the Crab Nebula (Becker \& Helfand
1983; Becker, Helfand \& Szymkowiak 1983; Blanton \& Helfand 1996).
It has long been suspected that this bright core component of Kes~75
might harbor a young pulsar, given its strong radio polarization and
flat spectral index, but none had been detected in any wave-band
(Becker, Helfand \& Szymkowiak 1983).  We have observed the region
around this supernova remnant several times with the X-ray detectors
aboard the Rossi X-ray Timing Explorer (\xte), while targeting a
nearby anomalous X-ray pulsar, and identified coherent pulsations at a
period of 0.32456 s. Since \xte\ is a non-imaging instrument only
crude directional information was available. Subsequent examination of
archival data from the Advanced Satellite for Cosmology and
Astrophysics (\asca) allowed us to locate the pulsed emission to the
core of the Kes 75 remnant. With five period measurements spanning
seven years, we are able to derive a stable period derivative. The
analysis of these data and the results they imply are the subject of
this Letter.

\section{Observation and Analysis}

The main instrument on-board the \xte\ observatory (Bradt, Rothschild
\& Swank 1993) consists of five co-aligned collimated detectors known
collectively as the Proportional Counter Array (PCA; Jahoda et
al. 1996). The effective area of the combined detector is about $6500$
cm$^{2}$ at $10$ keV with a roughly circular aperture response of
$\sim 1^{\circ}$ FWHM. Moderate spectral information is available in
the $2-60$ keV energy band with a resolution of $\sim 16\%$ at 6
keV. The absolute timing uncertainty is $\sim 100$ $\mu$s (Rots et
al. 1998). The arrival time of each photon was corrected to the Solar
System barycenter using the JPL DE200 ephemeris. An Observation Log
giving the net observing times is presented in Table 1.

The discovery of the $0.3$~s pulsation was made using $2^{22}$-bin
Fast Fourier transforms on our Jan 2000 data set (see below). This
produced a significant coherent signal at $\simeq 0.32456$ s. No
higher harmonics were found.  We next folded the timeseries into
20-bin periodograms centered near 0.3246 s; a large deviation from a
uniform model was found at a period of $P = 0.3243870$ s ($\chi^2 =
220$; April 1999), $P = 0.3245634$ s ($\chi^2 = 300$; Jan 2000), and
$P = 0.3246446$ s ($\chi^2 = 700$; June 2000).  With these three
observations, we were able to interpolate a preliminary period
derivative of $\dot{P}= 7.1 \times 10^{-12}\ {\rm s \ s^{-1}}$.

We then examined archival data of the region acquired by the \asca\
observatory (Tanaka, Inoue \& Holt 1994). Two observations were
available with sufficient duration and time resolution for a pulsar
search; of these, one is a deep 45~ks \asca\ pointing centered on
\snr\ whose analysis is detailed in Blanton \& Helfand 1996. We
searched data obtained with the Gas Imaging Spectrometer (GIS) for
evidence of pulsed emission around the period extrapolated from the
\xte\ observations assuming a constant period derivative. A total of
$12,800$ photons from the two GISs were extracted from a $8^{\prime}$
diameter aperture restricted to the hard energy range of $3-10$
keV. The selection of the hard energies pre-filters the softer
background emission from the SNR. Photon arrival times were again
corrected to the Solar System barycenter and then folded into
periodograms in order to search for a coherent signal. A faint but
unmistakable signal ($\chi^2 = 70$ for 10 phase bins) was found at the
expected period. This result was reproduced ($\chi^2 = 40$ for 10
phase bins) in the second \asca\ observation of 28 Mar 1999 which
contained the SNR in the field-of-view, $22^\prime$ off-axis.

\bigskip
\centerline{\psfig{figure=kes75_pulsar_fig1.ps,height=3.5in,angle=270,clip=}} 
\noindent {\footnotesize {\bf Figure 1} ---
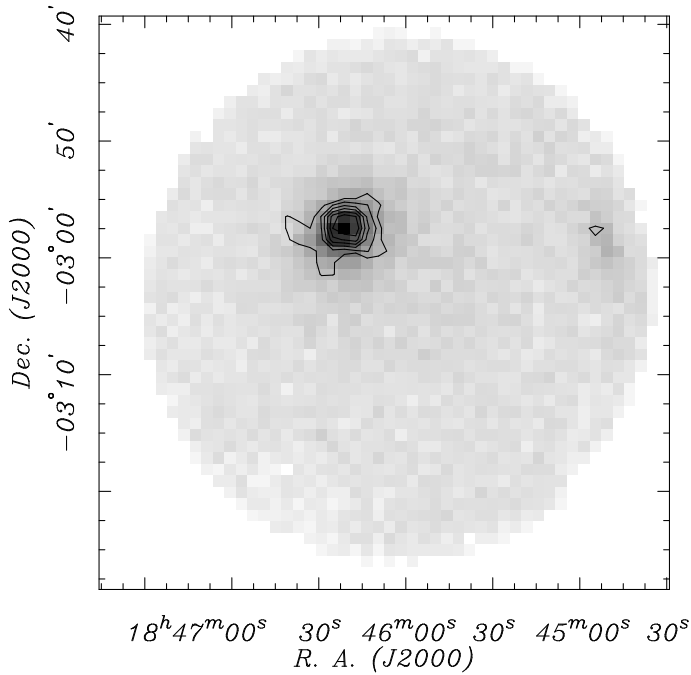
The \asca\ broad-band X-ray image (greyscale) of the region containing
the supernova remnant Kes 75.  The overlayed contour plot shows the
pulsed emission component from \psr, locating the pulsar to the
core of Kes 75.  The images are scaled by the square root of the intensity
while the contours are spaced in 10\% increments. 
}
\bigskip

To demonstrate that these pulsations are uniquely associated with
\snr, we performed phase-resolved image analysis as described in
Gotthelf \& Wang (2000). We generated images of the pulsed emission by
subtracting the off-pulse data from the on-pulse image. Only a single
source of significant emission remained: the pulsed emission, an
unresolved \asca\ point-like source located at the coordinates of the
SNR \snr\ (see Figure 1). We thus unambiguously locate the origin of
the pulsed emission to within an arcminute of the center of \snr.

Next, we used the \asca\ imaging data from the Solid-state Imaging
Spectrometer (SIS) to further refine the source location within the
SNR.  The X-ray emission from Kes 75 is known to have two components
(Blanton \& Helfand 1996), i.e. softer thermal emission of the shocked
interstellar medium of the SNR shell which cuts off above 3 keV, and a
hard component with a power-law spectrum which is unresolved with
\asca\ imaging and is located within the boundaries of the SNR. This
hard source, presumably due to non-thermal emission by confined
relativistic particles, was thought to contain the pulsar and possibly
an extended plerionic nebula.  The SIS coordinates of the hard point
source, the putative pulsar, are R.A. $18^h 46^m 24.5^s,
Dec. -02^{\circ} 58^{\prime} 28^{\prime\prime}$ (J2000) with an
uncertainty of $12^{\prime\prime}$ after correcting for the \asca\
temperature variation by the method of Gotthelf et al. (2000) This
position lies within $10^{\prime\prime}$ of the center of the
Crab-like emission in the middle of the radio remnant (Becker, Helfand
\& Szymkowiak 1983).

Employing the new localization, we re-barycentered all 5 observations
and re-determined the pulse periods using the trial period
derivative. A least-square fit of these period measurements to a
constant spin-down model gives an ephemeris for the Kes 75 pulsar with
period $ P = 0.32359795 \pm 0.00000012$ s and period derivative $ \dot
P = 7.09706 \pm 0.00094 \times 10^{-12}$ s/s, referenced to epoch MJD
$50000.0$.  We also fitted the data points to a spin-down model with a
second period derivative term.  Although the fits to both models were
found to be statistically acceptable with $ \chi_{\nu}^2 \simlt 1$, it
was not improved by adding the second derivative.  Figure 2 shows the
period evolution and the residuals from the best-fit linear model; our
estimates of the period uncertainty for each data point in Figure 2
are derived by using the method of Leahy (1983). We find no evidence
of period glitches in our measurements, as are frequently observed in
other young rotation-powered pulsars; however, considering the sparse
sampling they are by no means ruled out.  A summary of the timing
results for each data set used in this study is given in Table 1.

\bigskip
\centerline{\psfig{figure=kes75_pulsar_fig2.ps,height=2.5in,angle=270,clip=}} 
\noindent {\footnotesize {\bf Figure 2} ---
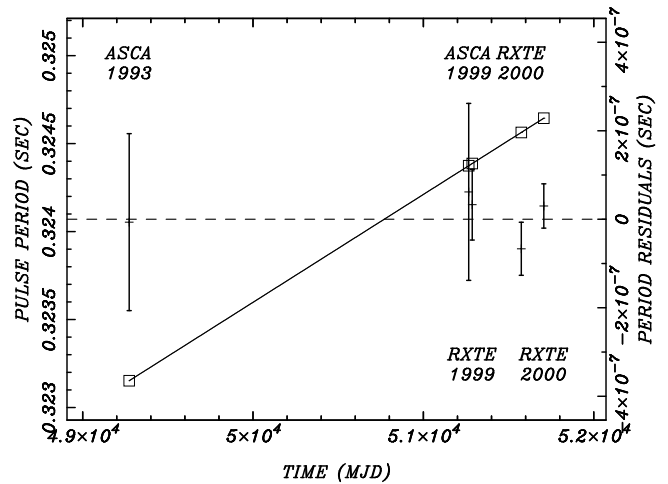
The pulse period evolution of \psr. (Left) Displayed are the period
measurements along with the best fit to a linear spin-down model (see
text). (Right) The residuals from this fit are shown along with their
$1$-$\sigma$ uncertainty error bars computed using the method of
Leahy (1987). The pulsar is evidently undergoing steady spin-down
at a rapid rate with no significant timing noise or noticeable glitch
activity.}
\bigskip

Using the above ephemeris, we generated pulse profiles for the \asca\
and the \xte\ combined data sets. These profiles are displayed in
Figure 3. They are well characterized by a single, broad peaked pulse
with a $\sim 50\%$ duty cycle. The profiles appear similar. The pulsed
emission comprises $5.5\%$ of the total \asca\ counts in the
background-subtracted folded light curves, suggesting that steady
emission from a synchrotron nebula might be dominating the flux.  This
is the same fraction as measured for the Crab pulsar (Seward 1984).

The \asca\ spectrum of Kes 75 with its hard non-thermal component is
presented in Blanton \& Helfand (2000). We have analyzed the \xte\
spectrum of the system above $3$ keV so as to exclude remnant's
thermal emission. The $3 - 20$ keV \xte\ spectrum was fitted to an
absorbed power law model with an absorbing column of N$_H = 3.1 \times
10^{22}$ cm$^{-2}$, the \asca\ value, and used the standard background
model along with a Gaussian line at 6.5 keV corresponding to the
Galactic Ridge diffuse Fe emission. We obtained a good fit for a
photon index $\Gamma = 2.18 \pm\ 0.04$ ($\chi_{\nu}^2 = 1.4$), typical
of a young, Crab-like pulsar and consistent with the \asca\ value.
The unabsorbed flux from the pulsar plus synchrotron nebula emission
is $1.8 \pm\ 0.4 \times 10^{-11}$ erg cm$^{-2}$ s$^{-1}$.  Next, we
considered the spectrum of the pulsed emission alone by analyzing
phase dependent spectra. Using the off-pulse spectrum as background,
we measure a hard photon index $\Gamma = 1.1 \pm\ 0.3$ ($\chi_{\nu}^2
= 1.5$) with an unabsorbed flux of the pulsed component of $0.96 \pm\
0.2 \times 10^{-12}$ erg cm$^{-2}$ s$^{-1}$ in the $3-10$ keV energy
band.

\bigskip
\centerline{\psfig{figure=kes75_pulsar_fig3.ps,height=2.5in,angle=270,clip=}} 
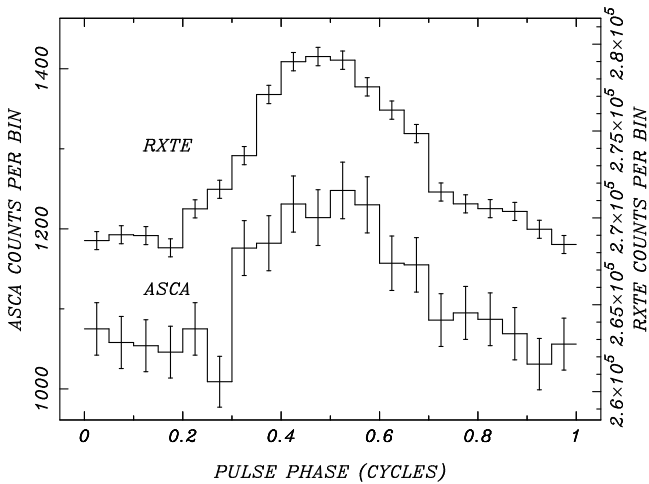
\noindent {\footnotesize {\bf Figure 3} --- The pulse profile of young
pulsar \psr\ in the $3-10$ keV band from the \xte\ (top) and \asca\
(bottom) observations folded into 20 phase bins using the ephemeris
presented in the text. The two profiles have been aligned to place the
peaks at the 0.5 phase bin; the relative phases are arbitrary. The
profiles include 105~ks and 93~ks of \xte\ and \asca\ data,
respectively.  }
\bigskip

\section{Discussion}

The steady increase in the period suggests a constant energy loss at a
rate of ${\dot E} = (2\pi)^2 I{\dot P}/P^{3} \simeq 10^{37} I {\rm \
ergs \ s^{-1}}$, where $I$ is the NS moment of inertia in units of
$1.4 \times 10^{45}$ g cm$^{2}$. In the standard pulsar model of a
rotating magnetized dipole, spin-down energy is lost via magnetic
dipole radiation, ${\dot E} \sim (B_p^2 R^6 \Omega^4)/(6c^3)$. This
model assumes a braking index of 3 and a uniformly magnetized stellar
interior. From the above relationships we can infer the surface
magnetic field at the pole, which, for \psr, gives $B_p \simeq
4.8\times 10^{13}$ G. Here, $R\sim 10{\rm\ km}$ is the neutron star
radius, $\Omega=2\pi/P$ is the angular velocity of the rotation, and
$c$ is the speed of light {\it in vacuo}. The characteristic spindown
age ($\tau=P/{2\dot{P}}$) of \psr\ is $\simeq 723$ yr, the youngest
characteristic age of any known pulsar.

For a well-behaved Crab-like pulsar with a braking index similar to 3,
the spindown age is reasonably consistent with that reported for
Kes~75 itself. Blanton \& Helfand (1996) estimated the age of the
remnant to lie between 900 and $4,300$ years, based on free expansion
and Sedov phase estimates, respectively. They preferred a value closer
to $10^3$ years, while arguing that the remnant was still in free
expansion phase, giving an age $\tau \sim 900 d_{19}v_5^{-1}$
years. Here, the distance to the SNR is $19d_{19}$ kpc (Milne 1979)
and the free expansion velocity is $5000v_5$ km s$^{-1}$.  Given the
quality of age estimators, the two limits are consistent. Finally, we
note that the cumulative energy in particles and fields within the
plerion -- a calorimetric measure of the pulsar's activity -- is
estimated from radio observations to be $10^{48}d_{19}^{17/7}$ erg
(Blanton \& Helfand 1996). This is also consistent with the lower
limit derived from the age of the pulsar and its current spin-down
loss, $\tau \dot E \sim 3\times 10^{47}$ erg.  From the spin
properties and location of \psr\ we infer an extremely young and
highly magnetized pulsar associated with the SNR Kes 75.

At the assumed distance to the SNR, the total X-ray luminosity from
the pulsar plus synchrotron nebula is $7.8 \times 10^{35}d_{19}^2$ erg
s$^{-1}$ in the $3-10$ keV band; for comparison, this translates to
$2.1 \times 10^{36}$ erg s$^{-1}$ in the ROSAT $0.1-2.4$ keV band for
the spectral parameters given above. The X-ray luminosity of this
region was already found to be the among the highest of any of the
Crab-like SNRs, and second only to the Crab itself, depending on the
true distance.  The derived pulsed luminosity of $4.1 \times 10^{34}
d_{19}^2$ erg s$^{-1}$ ($3-10$ keV) suggests that $\sim 4 \times
10^{-3}$ of the pulsars spin-down energy is emitted as pulsed X-rays
in the $3-10$ keV band or $\sim 1 \times 10^{-3}$ in the $0.1-2.4$
keV. This value is similar to those observed from other rotation
powered pulsars (Becker \& Tr\"umper 1999) and suggests that particle
acceleration is occurring in the NS magnetosphere. However, the ratio
of the total pulsar plus nebula luminosity to the spin-down energy
loss is 6$d_{19}^2$ times greater than that for the Crab pulsar.

In many ways, \psr\ resembles any other rotation-powered Crab-like
pulsar; however, its period, spin-down rate, spin-down conversion
efficiency, and inferred magnetic field, are each an
order-of-magnitude greater. The timing parameters of the new pulsar
are most similar to the young $0.4$-s pulsar J1119$-$6127 with its
period derivative of $4.0 \times 10^{-12}$ s/s implying $B_p = 4.1
\times 10^{13}$ G. This pulsar, though 4 times closer, does not
contain a similar bright radio or X-ray plerion (Camilo et al. 2000).

Recent research suggests that young neutron stars may have at least
two distinct evolutionary branches (see Gotthelf \& Vasisht 2000 and
refs. therein). Besides the Crab-like pulsars which evolve through
magnetic braking, with fields in the range $10^{12} - 10^{13}$ G and
spin periods at birth of order 10 ms, a second branch is made up of
the anomalous X-ray pulsars (AXPs; Mereghetti \& Stella 1998 and
refs. therein; Duncan \& Thompson 1996) and the soft $\gamma$-ray
repeaters (SGRs; Cline et al. 1982; Kulkarni \& Frail 1993), likely
``magnetars'', with magnetic fields in the range $10^{14} - 10^{15}$ G
(Vasisht \& Gotthelf 1997; Kouveliotou et al. 1998). The magnetars
typically have long spin periods, and their steady emission has only
been observed in the X-ray band. The new pulsar \psr\ lies in a
transitional regime with $B_p \simeq 5 \times 10^{13}$ G.

Few regular pulsars have implied magnetic fields strictly above the
quantum critical field, $B_{cr} \simeq 4.4 \times 10^{13}$ G; in fact,
the only such pulsar known thus far is the radio pulsar PSR
J1814$-$1744 which has an inferred field of $B_{13} \simeq 5.5$
(Pivavoroff, Kaspi \& Camilo 2000; Camilo et al. 2000), just above
this limit. Free electrons gyrate relativistically in $B > B_{cr}$
with radii less than the electron Compton wavelength, $\hbar/m_ec$. In
regular pulsars such as the Crab and Vela the purely quantum process
of single photon pair-production $\gamma \rightarrow e^{+}e^{-}$ is
invoked as a source of particle acceleration (Sturrock 1971). It has
been suggested that for magnetars with $B > B_{cr}$, the quantum
electrodynamical process of photon splitting ($\gamma \rightarrow
\gamma\gamma$), may compete with pair production and act as a
quenching mechanism for electrons, and suppress radio emission (Baring
\& Harding 1998). Note that the large pulse duty cycle in \psr\
suggests that unlike the Crab, which has a sharp high-energy pulse,
the X-ray producing particles in this pulsar are largely in the outer
magnetosphere.  Also, it remains to be seen if \psr\ is detectable as
a radio pulsar.  The shape of the radio pulse profile, vis-a-vis the
X-ray pulse would also be of considerable interest.

\begin{acknowledgements}
{\noindent \bf Acknowledgments} --- We thank J. P. Halpern and
F. Camilo for a careful reading of the manuscript and helpful
comments. This work is supported by NASA LTSA grant NAG5--7935.  This
research has made use of data obtained through the High Energy
Astrophysics Science Archive Research Center Online Service, provided
by the NASA/Goddard Space Flight Center.
\end{acknowledgements}

\begin{deluxetable}{lccccccc}
\tablecaption{Observation Log of Kes 75\vfill
\label{tbl-1}}
\tablehead{
 &\colhead{Date$^a$} & \colhead{Epoch} & \colhead{Exposure/} & \colhead{Period} & \colhead{Uncertainty} & \colhead{Offaxis$^b$}   & \colhead{Pulsed Flux$^c$} \cr
     &    & &   \colhead{Duration}       &   &    & & \colhead{$\times 10^{-12}$}\cr
  \colhead{Mission}      &   \colhead{(UT)}     &  \colhead{(MJD)}&   \colhead{(ks)}       &  \colhead{(s)}  &     \colhead{(ns)}    & \colhead{(arcmin)} & \colhead{(ergs cm$^{-2}$ s$^{-1}$)}\cr
}
\startdata
\asca \dots & 1993 Oct 10 & 49273.045014 & 44.5/86  & 0.32315219  & 200.0  & 7 & $1.2$ \cr
            & 1999 Mar 28 & 51265.924020 & 48.7/140 & 0.32437427  & 100.0  & 20&  \dots\ \cr
\xte \dots  & 1999 Apr 18 & 51286.667628 & 27.6/120 & 0.324386957 & 80.0   & 23& $0.6$ \cr
            & 2000 Jan 30 & 51574.527481 & 39.6/140 & 0.324563369 & 60.0   & 23& $ 0.8$ \cr
            & 2000 Jun 11 & 51706.829628 & 37.7/88  & 0.324644591 & 50.0   & 0.1& $1.0$ \cr
\enddata
\tablenotetext{a}{For the start of the observation.}
\tablenotetext{b}{Distance of Kes 75 from the obsevatory pointing direction.}
\tablenotetext{c}{The unabsorbed pulsed \xte\ and \asca\ flux using an absorbed power law model of photon index 1.1, in the $3 - 10$ keV energy band. The \asca\ GIS flux is derived using photons extracted from $4^{\prime}$ radius aperture centered on the \asca\ \snr\ position. The off-pulse spectrum was used as a background. See text for details.}
\end{deluxetable}

\end{document}